# A Point-to-Distribution Joint Geometry and Color Metric for Point Cloud Quality Assessment

Alireza Javaheri, Catarina Brites, Fernando Pereira, and João Ascenso
*Instituto de Telecomunicações - Instituto Superior Técnico*
Lisbon, Portugal
{alireza.javaheri, catarina.brites, fp, joao.ascenso}@lx.it.pt

*Abstract*—Point clouds (PCs) are a powerful 3D visual representation paradigm for many emerging application domains, especially virtual and augmented reality, and autonomous vehicles. However, the large amount of PC data required for highly immersive and realistic experiences requires the availability of efficient, lossy PC coding solutions are critical. Recently, two MPEG PC coding standards have been developed to address the relevant application requirements and further developments are expected in the future. In this context, the assessment of PC quality, notably for decoded PCs, is critical and asks for the design of efficient objective PC quality metrics. In this paper, a novel point-to-distribution metric is proposed for PC quality assessment considering both the geometry and texture. This new quality metric exploits the scale-invariance property of the Mahalanobis distance to assess first the geometry and color point-to-distribution distortions, which are after fused to obtain a joint geometry and color quality metric. The proposed quality metric significantly outperforms the best PC quality assessment metrics in the literature.

## I. Introduction

The recent advances in visual acquisition and visualization, i.e., sensors, cameras and displays, have boosted the development of more powerful visual representation models and formats, offering highly immersive and realistic visual experiences. The most popular of these representation models are the so-called *light fields,* which represent the visual scene as a dense array of views from multiple perspectives, and *point clouds* (PCs), which represent the visual scene through a large set of points (with attributes, e.g., color) in the objects' surfaces. The PC representation model is especially powerful to offer highly immersive, interactive, and realistic experiences, allowing the users to navigate a virtual world with the so-called *6 degree-of-freedom* (6-DoF) [1].

This 3D visual model matches the needs of application domains such as virtual and augmented reality, immersive communications, and autonomous driving [1]. However, a high-quality PC involves a large number of points, and thus visual data, which is challenging to transmit and store without compression. This means that large deployment of PC-based applications critically asks for efficient PC coding solutions and standards to offer interoperability among all these applications. As for images and video in the past, lossy PC coding solutions are used to lower the bitrate required for its transmission, commonly at the cost of introducing some distortion artifacts that may lower the users' perceived quality. Recently, MPEG has developed two PC coding standards, notably Geometry-based Point Cloud Compression (G-PCC) [3] for static and progressive PCs, and Video-based Point Cloud Compression (V-PCC) [4] for dynamic PCs. Naturally, along their development, it has been critical to assess the quality of the decoded PCs to evaluate and optimize the standards' compression performance.

Acknowledging the growing relevance of PC-based content in multiple application domains, the development of PC objective quality assessment metrics has become essential not only to design and optimize new PC coding solutions but also to assess and validate the quality of the PC content made available to users, in many application scenarios. While several PC objective quality metrics are already available, e.g., the D1-PSNR and D2-PSNR PC geometry quality metrics [5] used in MPEG and JPEG, they exhibit performance limitations. These metrics follow a simple distance-based approach, where the distance between corresponding points in the degraded and reference PCs is the key distortion error, thus ignoring the context and potential masking effects. Depending on the way these correspondences are made, point-to-point (Po2Po as D1-PSNR) and point-to-plane (Po2Pl as D2-PSNR) metrics are obtained.

To overcome the weaknesses of the currently available distance-based quality metrics, the authors have proposed a full reference, so-called *point-to-distribution (P2D)* quality metric for PC geometry [6], which adopts a new type of correspondence between two PCs, namely between a point in one PC and a distribution of points from a small neighborhood in the other PC. The idea underpinning this class of PC quality metrics is to statistically characterize the PC local surface, thus taking into account some spatial context and not only single points. This type of quality metric is not influenced by the ratio between the number of points in the decoded and reference PCs; this ratio has a strong impact, e.g., on the D1-PSNR metric accuracy, notably when the number of decoded points is much larger than the number of original points. The P2D *Mahalanobis* distance-based PC quality metric proposed in [6] has shown promising results for PC geometry quality assessment.

In this paper, P2D metrics are extended to evaluate PC color distortions. Moreover, using the scale-invariance property of the *Mahalanobis* distance, the P2D PC geometry and color distortion metrics are fused to obtain joint distortion and quality metrics with increased objective-to-subjective correlation performance. In this paper, the term 'distortion metric' refers to a metric which the scores grow with distortion (thus inversely to quality), and the term 'quality metric' refers to a metric in which scores grow with the quality (thus inversely to distortion). In summary, the main technical contributions of this paper are: i) novel P2D distortion and quality metrics for PC color; and ii) fused/joint P2D distortion and quality metrics for PC geometry and color, using the *Mahalanobis* distance scale-invariance property to evaluate the geometry and color distortions together.

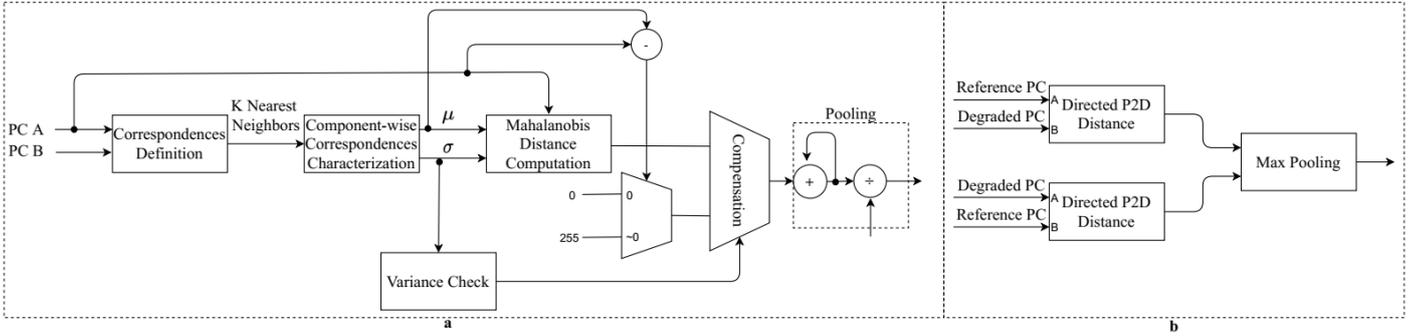

Fig. 1 P2D *Mahalanobis* distance-based color distortion/quality metric architecture: a) directed (PC A to PC B) P2D distance; b) symmetric P2D distance.

The rest of this paper is organized as follows. Section II briefly reviews the most recent and relevant works on PC objective quality assessment. Section III presents the proposed P2D joint geometry and color distortion and quality metrics in detail. Next Section IV presents the experimental results and their analysis. Finally, Section V concludes this paper with some final remarks.

## II. RELATED WORK

This section reviews the most recent PC objective quality metrics organized as point-based, feature-based, and projection-based metrics, depending on their approach.

The most popular point-based quality metrics for PC geometry are the point-to-point (Po2Po) [7] and point-to-plane (Po2Pl) [8] metrics, notably the D1-PSNR and D2-PSNR metrics largely used by MPEG and JPEG. In [9], Alexiou *et al.* propose the so-called *Plane-to-Plane (Pl2Pl)* metric, which measures the similarity between the normals associated to the plane's tangent to the surfaces for corresponding points in the reference and degraded PCs. Another approach is the so-called *Point-to-Surface (or Point-to-Mesh)* [10] which creates a polygonal mesh for the reference PC and measures the distance between each decoded point to the surface at the corresponding reference point, thus considering the underlying surface.

In [11], Javaheri *et al.* propose a geometry point-based quality metric based on the generalized Hausdorff distance. In [6], the same authors propose a P2D metric based on the *Mahalanobis* distance between a point and its $K$ nearest neighbors in the other PC. In [12], Javaheri *et al.* improve the popular D1-PSNR and D2-PSNR metrics by considering the PC rendering/intrinsic resolution as a normalization factor, thus obtaining a resolution adaptive PSNR metric. While there are not many point-based quality metrics for color, the Po2Po YUV-PSNR in the *YUV* color space is widely used by MPEG and in the literature to evaluate PC texture degradations. This metric may be computed only for the luminance (Y-PSNR) or chroma components (U/V-PSNR) individually, or as a weighted average of the three components as in YUV-PSNR.

Regarding the feature-based PC quality metrics, they typically consider both the geometry and color together. In [13], Meynet *et al.* propose the PC-MSDM metric, a structural similarity-based PC geometry quality metric based on local curvature statistics. In [14], Viola *et al.* design a PC quality metric based on the histogram and correlogram of the luminance component. In [15], Diniz *et al.* propose the so-called *Geotex* metric based on Local Binary Pattern (LBP) descriptors adapted to PCs and applied to the luminance. In [16], Diniz *et al.* extend the Geotex metric by considering multiple distances, notably Po2Pl MSE for geometry and the distance between LBP statistics from [15]; the quality score is the linear combination of the Po2Pl MSE distance between points and their nearest neighbor in the other PC, Po2Po MSE distance between LBP feature maps, and the distance between the LBP feature map histograms, with the weights for this linear combination obtained using a least-square error optimization. In [17], Diniz *et al.* propose a quality metric computing Local Luminance Patterns (LLP) on the $K$-nearest neighbors of each point on the other PC. In [18], Meynet *et al.* propose the Point Cloud Quality Metric (PCQM) metric, which combines three geometry features related to curvature with five color features related to lightness, chroma, and hue. In [19], Viola *et al.* propose a reduced reference quality metric jointly evaluating the geometry and color quality. Inspired by the SSIM quality metric for 2D images, Alexiou *et al.* propose in [20] the PointSSIM quality metric using statistical local dispersion features for geometry, color (luminance), normal and curvature information.

Projection-based quality metrics map PC 3D points on some 2D planes and evaluate the 3D quality using conventional 2D image quality metrics. In [21], Queiroz *et al.* propose a metric projecting the PC onto the six faces of a cube; after the six projected images are concatenated and the PSNR computed. In [22], Alexiou *et al.* design a rendering software for PC visualization on 2D screens, and the projected images are used for quality assessment with 2D image quality metrics. In [23], Alexiou *et al.* study the impact of the number of projected 2D images (each for a specific view) on the objective-to-subjective correlation performance of projection-based quality metrics. In [24], the same authors, use the metric in [22] to study the impact of using a different number of views, different pooling functions, etc. In [25], a 2D quality metric for texture and depth maps is proposed to assess the PC quality; RGB to Gaussian Color Model (GCM) conversion is performed to use a color space more aligned with the human visual system. The depth-edge map is aggregated upon the texture similarity between the reference and degraded PC as a local feature.

## III. PROPOSED P2D METRIC FOR JOINT GEOMETRY AND COLOR PC QUALITY ASSESSMENT

The P2D approach for PC distortion/quality assessment is first proposed in [6] targeting the assessment of PC geometry distortions by computing the distance between a point on the reference (or degraded) PC and the distribution of points in a small neighborhood around the corresponding point in the degraded (or reference) PC; in this way, the context around specific points is considered and not only a single corresponding point, which may be an outlier, thus also allowing to account for masking effects.

P2D distortion and quality metrics have the following key characteristics: 1) *Scale invariance*, i.e., can evaluate PCs with different intrinsic characteristics in terms of precision (i.e., coordinates bit-depth) and resolution (i.e., the average distance between neighboring points); 2) *Correlation awareness*, i.e., consider the correlation between points and their neighbors and weight the distances in each coordinate accordingly; naturally, points not following the surface distribution will lead to larger distances; 3) *Normals computation free*, i.e., do not require the estimation of normals to model the underlying surface, which is a great feature as normal computation is an error-prone and computationally complex process, e.g. used in Po2Pl and Pl2PL metrics.

In this paper, a P2D distortion metric for PC color is proposed, inspired by the P2D metric for PC geometry of [6]. The two distortion and color metrics (for geometry and color) between the degraded and reference PCs are after fused, exploiting the scale invariance property of the *Mahalanobis* distance. For each distortion metric, the corresponding quality metric is proposed with the same MOS estimation accuracy as the distortion metric.

*A. P2D Mahalanobis-based Color Distortion Metric*

The P2D color distortion between the degraded and reference PCs is computed for each color component independently and after fused together. As usual in the literature, the YUV color space will be adopted here, thus acknowledging the high impact of luminance on visual quality assessment. Since color components are linear independent [26], this is sufficient to assume they are uncorrelated. By considering no correlation between the color components, their covariance matrix is diagonal and the *Mahalanobis* distance is reduced to the Standardized Euclidean distance, which is equivalent to the regular Euclidean distance after data normalization based on the mean and standard deviation.

Fig. 1 (a) shows the architecture for the proposed P2D *Mahalanobis*-based directed (degraded to reference PC or vice-versa) color distance/distortion while Fig. 1(b) shows the architecture for the corresponding P2D symmetric distance/distortion. As for geometry [6], this P2D distance is computed in both directions, i.e., degraded to reference (e.g., from PC $A$ to $B$) and vice-versa. For point $\vec{a} = (x, y, z)^T$ in PC $A$, with color $\vec{C_a} = (Y, U, V)^T$ the directed P2D distance for color is computed for each color component, $C_m$, independently as:

1. **Correspondences Definition:** Find the set of points $D_a$ which are the $K$ nearest neighbors of $\vec{a}$ in PC $B$. This local neighborhood represents the support region for the P2D metric in PC $B$. Naturally, the $K$ value is rather important as it defines the size of the local neighborhood and thus the size of the context; when $K$ is too small, spatial correlation between points may not be fully captured and, when $K$ is too large, fine details may be missed.

2. **Component-wise Correspondence Characterization:** For each color component $C_m$, compute the mean and variance for $D_a$ as follows:

$$\mu_{Cm} = \frac{1}{K} \sum_{i=1}^{K} C_{i,C_m}, \quad (1)$$

$$\sigma_{C_m}^2 = \frac{1}{K} \sum_{i=1}^{K} (C_{i,C_m} - \mu_{C_m}), \quad (2)$$

where $C_{i,C_m}$ is the $Cm^{th}$ color component value for point $i$ and $K$ is the number of nearest neighbors in $D_a$. These parameters statistically characterize the distribution of points in the neighborhood $D_a$. Since there is no correlation between color components, $\sigma_{YCb}$, $\sigma_{YCr}$, $\sigma_{CbCr}$ are equal to zero.

3. **Mahalanobis Distance Computation:** When the correlation between random variables is zero, the *Mahalanobis* distance is equivalent to the Standardized Euclidean Distance. This distance for each point $a$ in PC $A$ to the neighborhood $D_a$ in PC $B$ is characterized by $\mu_{C_m}$ and $\sigma_{C_m}^2$, for color component $Cm$, comes as follows:

$$d_{a,B}^{P2D\text{-}C_m} = \sqrt{\frac{(C_{a,C_m} - \mu_{C_m})^2}{\sigma_{C_m}^2}}, \quad (3)$$

for example, for the luminance, this distance would become $d_{a,B}^{P2D\text{-}Y}$.

4. **Variance Check and Compensation:** If the variance of any color component is zero, (3) cannot be computed; in that case, the distance in (3) would be either infinity or ambiguous. Since a zero variance means that all the points in the neighborhood/distribution $D_a$ have similar color, if point $a$ shares the same value as the mean of all points in $D_a$, then the error should be zero. Otherwise, the distance should be the maximum possible distance, notably 255 for 8-bit color values.

5. **Pooling:** Finally, average pooling is used to aggregate the P2D distances for all points in the PC A:

$$d_{A,B}^{P2D\text{-}C_m} = \frac{1}{N_A} \sum_{a \in A} d_{a,B}^{P2D\text{-}C_m}. \quad (4)$$

As usual, the final (symmetric) P2D distance is pooled as the maximum of the two directed distances (degraded to reference PC and vice-versa), thus obtaining the P2D distortion metric for each color component:

$$P2D\text{-}C_m = \max(d_{A,B}^{P2D\text{-}C_m}, d_{B,A}^{P2D\text{-}C_m}). \quad (5)$$

If three color components are to be considered, a combined *P2D-YUV* may be defined using the usual weighting for the three color components as for *PSNR-YUV*:

$$P2D\text{-}YUV = \frac{6 \times P2D\text{-}Y + P2D\text{-}U + P2D\text{-}V}{8}. \quad (6)$$

*B. P2D Joint Geometry and Color Distortion and Quality Metrics*

P2D distortion metrics are unitless and scale-invariant as they are based on the distance between a point in PC *A* and the mean of the distribution for the corresponding *K*-nearest neighbors in PC *B* measured using the scale-invariant *Mahalanobis* distance.

The P2D distortion metric scale-invariance property allows that a fused/joint distortion metric is obtained by adding the distortion metric values for the color components and geometry, since they are in the same scale after Step 3 in the algorithm above. It was previously shown in [6] that the scale-invariant *Mahalanobis* distance makes possible to compare the quality of PCs with different geometry precisions.

Exploiting this property, a fused/joint geometry and color P2D distortion metric is proposed, considering three different pooling functions, notably min, max, and average as follows:

$$P2D_{min}\text{-}JGC_m = \min(P2D\text{-}C_m, P2D\text{-}G), \quad (7)$$

$$P2D_{max}\text{-}JGC_m = \max(P2D\text{-}C_m, P2D\text{-}G), \quad (8)$$

$$P2D_{avg}\text{-}JGC_m = \mathrm{avg}(P2D\text{-}C_m, P2D\text{-}G), \quad (9)$$

where $P2D\text{-}G$ is computed as defined in [6]. If only the luminance component is considered, the joint luminance and geometry distortion metric ($P2D\text{-}JGY$) with average pooling leads to:

$$P2D\text{-}JGY = \frac{P2D\text{-}Y + P2D\text{-}G}{2}. \quad (10)$$

Since a distance is measured, the P2D joint distortion metric scores grow with the distortion and inversely to the quality. If a P2D quality metric is preferred, with the metric scores growing with quality, a logarithmic variant of the P2D distortion metrics may be defined as follows:

$$LogP2D = \log_{10}\left(1 + \frac{1}{P2D}\right). \quad (11)$$

These LogP2D quality metrics are derived from the P2D distortion metrics by inverting the distortion score and adding 1 to avoid negative values. It is possible to apply (11), to any P2D quality metric, resulting in $LogP2D\text{-}G$, $LogP2D\text{-}Y$ and $LogP2D\text{-}JGY$. As will be shown in Section IV, the subjective-to-objective correlation performance is the same between the LogP2D and P2D distortion and quality metrics.

## IV. PERFORMANCE ASSESSMENT

This section intends to study and assess the proposed P2D joint geometry and color distortion and quality metrics, notably in comparison with state-of-the-art PC quality metrics using meaningful test conditions, i.e., a set of representative PCs and relevant PC codecs.

### A. Test Material and Conditions

To perform this assessment, the MPEG Point Cloud Compression Dataset (M-PCCD), created at EPFL and publicly available in [27], has been selected. This recent dataset includes 232 stimuli (a combination of a PC coded with a specific codec at a specific rate) where geometry and color are both coded; the dataset includes both the MOS values as well as the reference and decoded PCs. This is the type of dataset required to assess and validate a PC joint geometry and color distortion/quality metric as those proposed in this paper. The test material in this dataset corresponds to nine PCs, including four objects and five human figures, notably *Longdress*, *Loot*, *Redandblack*, *Soldier*, *The20smaria* and *Head* from MPEG repository [28], *Romanoillamp* and *Biplane* from JPEG repository [29] and *Amphoriskos* from *Sketchfab* [30]. *Redandback* is also used for subject training in this dataset.

The PCs have been coded in the following conditions: i) 24 rates for the MPEG G-PCC standard with six different rates for each combination of Octree and TriSoup geometry coding modes with the RAHT and Lifting color coding modes; and ii) five rates for the MPEG V-PCC standard. The rates were selected based on the MPEG Common Test Conditions (CTC) recommendations [5].

A non-linear regression has been used to map the computed P2D objective distortion/quality scores into the MOS scale; in this case, a logistic function has been adopted to obtain the predicted MOS scores as [31]:

$$MOS_p = \beta_2 + \frac{\beta_1 - \beta_2}{1 + e^{-\left(\frac{Q_i - \beta_3}{\beta_4}\right)}}, \quad (12)$$

where $Q_i$ is the objective metric values and $\beta_1, \ldots, \beta_4$ are the regression model parameters.

To assess the objective-to-subjective correlation for the metrics under study, the Pearson Linear Correlation Coefficient (PLCC) and Spearman Ranked Order Correlation Coefficient (SROCC) are used as common in the literature; for a more complete assessment, the Root Mean Squared Error (RMSE) is also provided.

In this section, the P2D distortion and quality metrics may be referred without distinction since their correlation performance is essentially the same.

### B. Best Color Component Assessment

The P2D color metric of (5) can measure the color distortions for any of the color components, notably Y, U and V as well as for YUV using (6). In this section, the correlation performance for the P2D metric using the most relevant color components is compared in Fig. 2, where both SROCC and PLCC in function of the neighborhood size are shown. From the charts, it is possible to conclude that the P2D luminance only metric performs significantly better than the other P2D color component metrics including the YUV weighted average metric. For this reason, the following experiments will adopt the luminance only P2D color metric. It is relevant to highlight that this conclusion has been similar for other metrics in the literature, e.g., the PointSSIM metric [20].

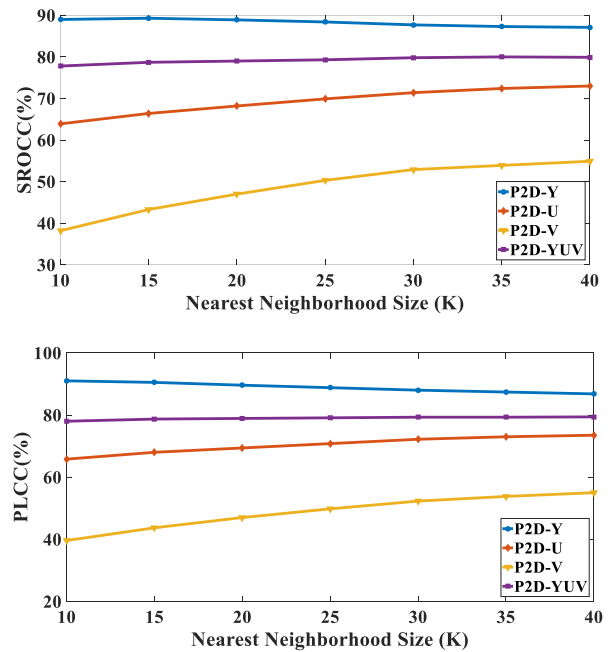

Fig. 2 SROCC (Top) and PLCC (Bottom) performance as a function of the nearest neighborhood size (K) for the P2D distortion metric using different color components.

### C. Best Neighborhood Size Assessment

The Correspondences Definition step in the P2D distortion computation requires the selection of an appropriate PC neighborhood, here characterized by the *K* parameter corresponding to the number of nearest neighbors considered. Since the *K* parameter has a major impact on the proposed metrics' performance, some experiments have been designed to study the P2D metrics performance as a function of *K*. The obtained PLCC and SROCC performances are presented in Fig. 3 and show that the impact of $K$ is not similar for the geometry and luminance.

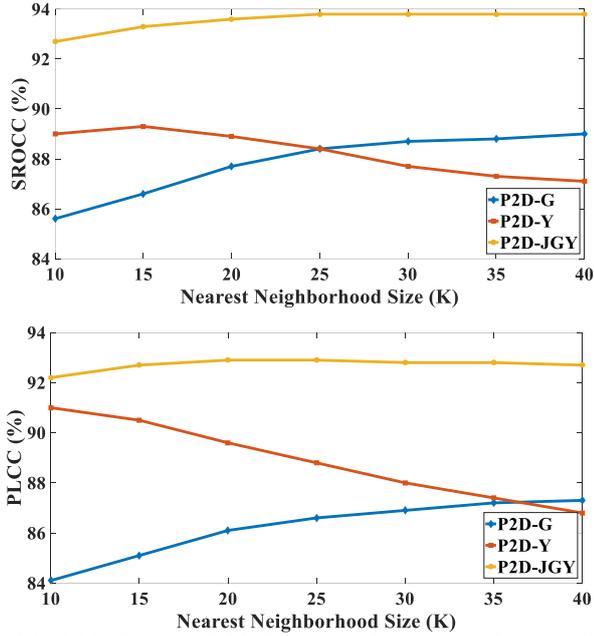

Fig. 3 SROCC (Top) and PLCC (Bottom) performance as a function of the nearest neighborhood size ($K$) for geometry, luminance, and joint P2D quality metric.

Fig. 3 shows that the P2D distortion metric correlation performance slightly increases with $K$ for the geometry while the opposite happens for the luminance. For the P2D joint metric, the best correlation performance occurs for the mid-range neighborhood size, since a very small decrease in PLCC can be observed for high values of $K$. In general, after choosing a large enough neighborhood size (larger than 10), the correlation performance does not change significantly by increasing the neighborhood size for the *P2D-JGY*. Following the experimental results, the selected neighborhood sizes for the *P2D-G*, *P2D-Y* and *P2D-JGY* joint distortion and quality metrics are 40, 15, and 25, respectively.

*D. Best Pooling Function Assessment*

The P2D geometry and luminance distortion metrics can be fused to obtain the P2D joint distortion metric by using one of the three pooling functions proposed in Section III.B (equations (7) – (9)). Fig. 4 shows the *P2D-JGY* correlation performance when using these pooling functions for different nearest neighborhood sizes ($K$). For most $K$ values, including the neighborhood sizes selected in the previous sub-section, the *average* pooling function is the best performing and thus will be used in the remaining experiments in this paper.

*E. P2D Distortion and Quality Metrics Performance*

Table I shows the objective-to-subjective correlation performance for many quality metrics from the literature and also those proposed in this paper, for each individual PC codec (G-PCC and V-PCC) and all codecs together (i.e., all data jointly considered), after logistic fitting. The following conclusions may be taken:

- *P2D distortion vs LogP2D quality metrics:* The correlation performance for the P2D distortion and P2D quality metrics is essentially the same; their use is only related to the preference of having directly or inversely increasing scores with the PC quality.

- *P2D Luminance vs Geometry vs Joint metrics*: The performance results show that the proposed *P2D-Y* performs better than the *P2D-G*, especially for the V-PCC codec with 14.1% and 12.3% higher SROCC and PLCC, respectively. When geometry and luminance are fused, the *P2D-JGY* outperforms *P2D-G* and *P2D-Y* with at least 4.5% and 2.2% higher SROCC and PLCC, respectively.

- *G-PCC vs V-PCC vs All codecs:* Performance gains for the P2D metric regarding state-of-the-art metrics are larger for the V-PCC codec than for the G-PCC codec and for All codecs. PLCC and SROCC performance for the proposed *P2D-JGY* joint metric is significantly better than for state-of-the-art metrics for V-PCC. For G-PCC, both the proposed *P2D-JGY* and PointSSIM perform very well.

- *Proposed P2D vs state-of-the-art quality metrics:* The proposed *P2D-JGY* joint metric outperforms all the state-of-the-art quality metrics when V-PCC and All codecs are considered; for G-PCC is at the top performance together with PointSSIM.

In summary, the proposed *P2D-JGY* (distortion or quality) metric is, overall, the best performing PC metric comparing to state-of-the-art metrics.

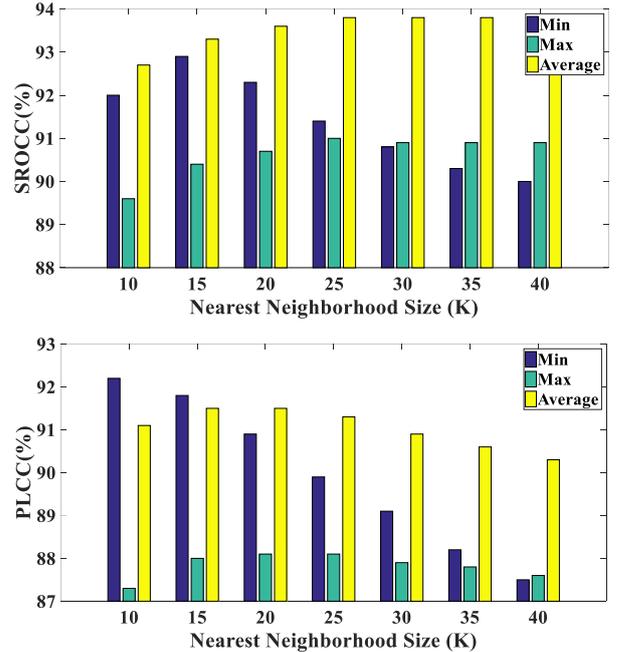

Fig. 4 SROCC (Top) and PLCC (Bottom) performance of *P2D-JGY* for the three pooling functions for varying neighborhood sizes, $K$.

## V. FINAL REMARKS

PC quality assessment is a critical technology for emerging immersive visual experiences based on this visual representation model. The joint quality assessment of color and geometry looks more appropriate since users see the two types of data together and impacting each other, e.g., through masking. This paper proposes a novel PC joint geometry and color quality metric based on the idea of characterizing a surface neighborhood through some statistical parameters and measuring the scale-invariant *Mahalanobis* distance. The correlation performance results show that the proposed P2D metric outperforms the available metrics in the literature.

TABLE I.
PLCC, SROCC, AND RMSE PERFORMANCE FOR THE PROPOSED P2D OBJECTIVE METRICS IN COMPARISON WITH STATE-OF-THE-ART METRICS

| Type | Metric | All Codecs Together | | | V-PCC Codec Only | | | G-PCC Codec Only | | |
|---|---|---|---|---|---|---|---|---|---|---|
| | | SROCC | PLCC | RMSE | SROCC | PLCC | RMSE | SROCC | PLCC | RMSE |
| Point-based (Po2Po) | D1-PSNR [7] | 79.7 | 77.7 | 0.857 | 28.2 | 30.4 | 0.997 | 83.9 | 82.5 | 0.794 |
| | GH 98% PSNR [11] | 86.9 | 84.6 | 0.726 | 57.3 | 57.8 | 0.854 | 89.9 | 88.5 | 0.653 |
| | RA-PSNR ($APD_{10}$) [12] | 90.2 | 88.8 | 0.626 | 67.3 | 68.9 | 0.759 | 91.8 | 91.0 | 0.584 |
| | Y-PSNR [7] | 66.2 | 67.1 | 1.009 | 33.3 | 37.6 | 0.970 | 70.3 | 71.4 | 0.984 |
| Point-based (Po2Pl) | D2-PSNR [8] | 83.8 | 80.5 | 0.808 | 55.3 | 60.3 | 0.835 | 87.3 | 83.4 | 0.774 |
| | GH 98% PSNR [11] | 87.9 | 84.3 | 0.731 | 71.2 | 75.1 | 0.691 | 91.0 | 87.5 | 0.680 |
| | RA-PSNR ($APD_{10}$) [12] | 89.9 | 88.9 | 0.622 | 76.9 | 79.9 | 0.629 | 91.5 | 90.3 | 0.604 |
| Feature-based | PointSSIM [20] | 91.8 | 92.6 | 0.514 | 84.5 | 83.0 | 0.584 | 92.9 | **94.4** | **0.462** |
| | $d_{gc}$ [14] | 92.0 | 90.4 | 0.585 | 74.0 | 75.3 | 0.689 | 93.9 | 92.5 | 0.533 |
| | $H^Y_{L2}$ [14] | 88.4 | 85.3 | 0.710 | 68.3 | 65.7 | 0.789 | 91.9 | 87.9 | 0.669 |
| | $PCM_{RR}(MCCV)$ [19] | 90.7 | 90.2 | 0.573 | 64.8 | 71.6 | 0.731 | 91.0 | 89.2 | 0.636 |
| Projection-based | Y-MS-SSIM [24] | 75.2 | 70.9 | 0.959 | 35.4 | 31.9 | 0.992 | 50.1 | 75.3 | 0.924 |
| | Y-VIFP [24] | 74.2 | 71.5 | 0.951 | 35.6 | 43.7 | 0.942 | 79.2 | 75.0 | 0.929 |
| Point-based (Proposed P2D and LogP2D) | P2D-G [6] (NN40) | 89.0 | 87.3 | 0.663 | 70.6 | 72.6 | 0.720 | 90.4 | 89.2 | 0.636 |
| | LogP2D-G [6] (NN40) | 89.0 | 87.3 | 0.664 | 70.6 | 72.7 | 0.719 | 90.4 | 89.1 | 0.639 |
| | P2D-Y (NN15) | 89.3 | 90.5 | 0.578 | 84.7 | 84.7 | 0.556 | 90.8 | 92.1 | 0.546 |
| | LogP2D-Y (NN15) | 89.3 | 90.7 | 0.574 | 84.7 | **85.0** | 0.552 | 90.8 | 92.3 | 0.541 |
| | P2D-JGY (NN25) | **93.8** | **92.9** | 0.503 | **85.5** | 84.2 | 0.564 | **94.6** | 94.0 | 0.477 |
| | LogP2D-JGY (NN25) | **93.8** | **92.9** | **0.502** | **85.5** | 84.3 | **0.563** | **94.6** | 94.1 | 0.477 |